\documentclass[11pt]{article}
\usepackage{epsfig} 
\newcommand{\sect}[1]{ \section{#1} \setcounter{equation}{0} } 

\newcommand{\half}{\mbox{\small{$\frac{1}{2}$}}}

\newcommand{\MSbar}{\overline{\mbox{MS}}} 

\newcommand{\Nf}{N_{\!f}}

\newcommand{\pslash}{p \! \! \! /}
\newcommand{\qslash}{q \! \! \! /}

\begin{document}
\title{Off-shell quark bilinear operator Green's functions at two loops}
\author{J.A. Gracey, \\ Theoretical Physics Division, \\ 
Department of Mathematical Sciences, \\ University of Liverpool, \\ P.O. Box 
147, \\ Liverpool, \\ L69 3BX, \\ United Kingdom.} 
\date{}

\maketitle 

\vspace{5cm} 
\noindent 
{\bf Abstract.} We construct the two loop Green's functions for a quark
bilinear operator inserted at non-zero momentum in a quark $2$-point function 
for the most general off-shell configuration. In particular we consider the
quark mass operator, vector and tensor currents as well as the second moment
of the flavour non-singlet Wilson operator. 

\vspace{-15.2cm}
\hspace{13.5cm}
{\bf LTH 1206}

\newpage 

\sect{Introduction.}

Recently an interesting study has appeared, \cite{1}, which concerns the mass 
composition of the proton using lattice gauge theory. It is now accepted that 
quarks and gluons are the fundamental constituent particles which form the 
hadrons. However the relative percentage contribution of each parton to the 
overall mass is not precisely known. In \cite{1} this breakdown was provided 
using lattice gauge theory methods and it was shown that around $9\%$ is 
attributable to the quark condensates from the weak sector of the Standard 
Model. Of the remainder quark and gluon energies contribute $32\%$ and $37\%$ 
respectively and the anomalous gluonic part makes up the remaining $23\%$. To 
achieve such results the underlying quantum field theory, Quantum
Chromodynamics (QCD), was used to study the energy-momentum tensor as well as 
other physically important operators. This is not a straightforward exercise 
since one has to operate in a strictly non-perturbative region of QCD,
\cite{1}. Moreover aside from estimating errors one has to ensure that the 
measurements and results are not inconsistent with known high energy behaviour.
By this we mean that the lattice computed Green's functions have to be 
consistent over all energy ranges. Therefore ensuring that measurements 
correctly extrapolate to the high energy limit is important. This was 
incorporated in the matching analysis of \cite{1} to high loop order 
computations in the chiral limit. However the early perturbative results of 
\cite{2,3,4,5} used in \cite{1} were for a specific external momentum 
configuration set up. For instance, the Green's functions used for the matching
correspond to a quark bilinear operator inserted in a quark $2$-point function.
In effect overall this becomes a $3$-point function since an external momentum 
can flow into the operator insertion in addition to those of the quark external 
legs. For the lattice matching used in \cite{1} the perturbative set up was the 
one where the operator momentum is zero, \cite{2}, and hence is an exceptional 
configuration. However, there is also interest for lattice computations in more
general configurations. For instance, in \cite{6,7,8} operators have been 
considered with a non-zero momentum insertion in the symmetric point 
configuration. This is known as the symmetric momentum (SMOM) case since the 
squares of each of the three external momenta are all equal. A similar 
configuration was used in \cite{9,10} for studying the $3$-point QCD vertex 
functions in the continuum. The set up of \cite{6,7,8} has proved to have had a
wide use in a variety of lattice problems involving quark bilinear operators. 
For instance, a non-exhaustive representation set of recent studies can be 
found in \cite{11,12,13,14,15,16,17,18}. 

Subsequently variations on this external momentum configuration scheme have
been considered, \cite{8,9} where the operator momentum squared differs from 
those of the external quarks which are equal to each other. One example of 
the usefulness of such set ups can be seen in \cite{14} where the 
renormalization constants of quark bilinear operators were computed in two 
different renormalization schemes on the lattice. One was the RI${}^\prime$ 
scheme of \cite{2,3} and the behaviour of those renormalization constants was 
compared to the corresponding ones in the SMOM set up. Interestingly for 
several operators the results for the latter scheme were reliable over a much 
larger energy range than the RI${}^\prime$ case. This was in the sense that in 
the chiral limit the mass operator and pseudoscalar operator renormalization 
constants should have the same value with a similar statement for the vector 
and axial vector currents in the flavour non-singlet case. That these agree for
virtually the whole energy range for the operator with non-zero momentum 
insertion in the Green's function provides credence to moving to the SMOM 
schemes for more reliable analyses. This may be due in part to them being
kinematic based schemes using non-exceptional momentum configurations where
infrared issues do not arise.  With the advances in lattice technology to allow
us to study internal hadron dynamics in more depth and precision there is a 
clear need for the continuum matching programme to progress too for quark 
bilinear operators as well as for other operators. One interesting recent
development on the experimental side is the measurement of the pressure exerted
by the constituent partons inside a hadron. For instance in \cite{19} the 
pressure distribution inside a proton was measured experimentally. Subsequently
there has been a lattice investigation into estimating the pressure 
distribution as well as shear forces inside the proton, \cite{20}. With the 
progress in the precise constituent mass breakdown of a proton in \cite{1} 
through operator measurements on the lattice, then to progress with theoretical
parton pressure studies will require lattice analysis too. This will also 
necessitate high loop results in the continuum field theory but for a more 
general momentum configuration than those such as SMOM used for the matching so
far. Therefore to keep apace of such developments it is the purpose of this 
article to extend the SMOM computations of the quark $2$-point functions with 
quark bilinear operator insertions to the most general off-shell momentum 
configuration. This will provide results for a large range of momentum transfer
cases including the one where all external momenta squared are different. In 
particular we will focus on the Green's function with flavour non-singlet 
scalar, vector and tensor operators inserted as well as the first moment of the
Wilson operator used in deep inelastic scattering. These will all be in the 
chiral limit. So we will not need to consider the axial vector of pseudoscalar 
operators. The various Green's functions will be computed to {\em two} loops in 
the $\MSbar$ scheme and we will provide the complete decomposition into the 
full basis of Lorentz tensors. This is important since it will allow in 
principle lattice measurements in a variety of different component directions. 
While the quark mass and vector current operators are standard quantities to 
consider, there has been interest in the tensor current in recent years, 
\cite{13,16,18}. For example, such operators can arise as part of dimension six
operators in effective field theory extensions of the Standard Model. In one 
recent study, \cite{16}, nucleon form factors of the tensor current have been 
examined with input from lattice QCD results. Another article recording a 
lattice study of tensor currents is \cite{13}. Therefore our off-shell 
computations will be useful for perturbative matching in future extensions of 
such lattice analyses.

The paper is organized as follows. We detail the quantum field theoretic
aspects of the machinery we will use in Section $2$ before recording our
results in Section $3$. Concluding remarks are given in Section $4$. Two
appendices are provided. The first records the tensor basis for the Green's
function of each operator considered together with the projection matrix. The
other summarizes the various analytic functions which appear in the one and two 
loop amplitudes.

\sect{Formalism.}

We outline the formal details of the various Green's functions we will
evaluate off-shell in this section and use parallel notation to previous
articles, \cite{21,22}. To assist with labelling of various quantities we will 
use the same shorthand notation for the following gauge invariant quark 
bilinear operators which is
\begin{equation}
S ~ \equiv ~ \bar{\psi} \psi ~~,~~
V ~ \equiv ~ \bar{\psi} \gamma^\mu \psi ~~,~~
T ~ \equiv ~ \bar{\psi} \sigma^{\mu\nu} \psi ~~,~~
W_2 ~ \equiv ~ {\cal S} \bar{\psi} \gamma^\mu D^\nu \psi ~~,~~
\partial W_2 ~ \equiv ~ {\cal S} \partial^\mu \left( \bar{\psi} \gamma^{\nu}
\psi \right)
\label{opset}
\end{equation}
where $\psi$ is the quark field and the gluon, $A^a_\mu$, is embedded in the
covariant derivative with coupling constant $g$. We note that all operators are
flavour non-singlet and $\sigma^{\mu\nu}$~$=$~$\half [\gamma^\mu,\gamma^\nu]$. 
Since we are concerned with the chiral limit then results for the pseudoscalar 
and axial vector operator will be the same as their respective scalar and 
vector counterparts and we will make no further reference to them. The two 
operators $W_2$ and $\partial W_2$ are symmetric and traceless with respect to 
their Lorentz indices in $d$-dimensions. We illustrate this by an example for 
the latter operator. Defining
\begin{equation}
{\cal O}^{\partial W_2}_{\mu\nu} ~=~ \partial_\mu \left( \bar{\psi} \gamma_\nu 
\psi \right)
\end{equation}
then 
\begin{equation}
{\cal S} {\cal O}^{\partial W_2}_{\mu\nu} ~=~ 
{\cal O}^{\partial W_2}_{\mu\nu} ~+~ {\cal O}^{\partial W_2}_{\nu\mu} ~-~ 
\frac{2}{d} \eta_{\mu\nu} {\cal O}^{\partial W_2\,\sigma}_{\sigma} 
\end{equation}
is the symmetric and traceless operator. Given the structure of the operator
$W_2$ one might expect that the operator where the covariant derivative acts
solely on the anti-quark is not included. However it is not an independent
operator since it can be written as a linear combination of $W_2$ and 
$\partial W_2$. We could have chosen to ignore the latter in place of a more
symmetric choice of independent operators. However one of the reasons we have
included $\partial W_2$ instead is that while it mixes under renormalization
with $W_2$, as would be the case for the other operator which we regard as not 
independent, the mixing matrix of the $\{W_2,\partial W_2\}$ set is triangular.
This produces a natural partition within the larger profile since the 
renormalization constant of $\partial W_2$ is the same as that of $V$. For the 
other operators $S$, $V$ and $T$ there is no mixing and their renormalization 
is purely multiplicative. For notational reasons if we label the sector 
containing both twist-$2$ Wilson operators by $W_2$ which should not lead to 
any ambiguity then the $W_2$ sector operators renormalize according to 
\begin{equation}
{\cal O}_{\mbox{\footnotesize{o}}\, i} ~=~ Z^{W_2}_{ij} {\cal O}_j
\end{equation}
where ${}_{\mbox{\footnotesize{o}}}$ indicates a bare entity. With our choice
of operator basis for the twist-$2$ operators the mixing matrix of 
renormalization constants will have the form  
\begin{equation}
Z^{W_2}_{ij} ~=~ \left(
\begin{array}{cc}
Z^{W_2}_{11} & Z^{W_2}_{12} \\
0 & Z^{W_2}_{22} \\
\end{array}
\right) ~.
\label{zmixdef}
\end{equation}
Another reason we have included the operator $\partial W_2$ is that it cannot
be neglected when one studies operator Green's functions where there is an
external momentum flowing into the operator. In the early work of 
\cite{23,24,25} the main interest was the renormalization of the Wilson 
operators themselves alone. The mixing with the total derivative operators was 
not needed. Therefore the operators were renormalized by inserting into a quark 
$2$-point function where there was {\em no} external momentum flow into the 
operator itself. In this set up there is no need to consider any mixing issues 
as the off-diagonal matrix element of (\ref{zmixdef}) could not be accessed and 
was not needed for the deep inelastic scattering formalism. As our motivation 
is to contribute to a different problem which involves knowing the structure of
a full Green's quark $2$-point function with an operator at non-zero momentum 
insertion the operator $\partial W_2$ must be included. By doing so we have a 
closed set of operators under renormalization. This has been tested in
\cite{26} where $2$-point operator correlation functions were computed to three
loops in the chiral limit for the set given in (\ref{opset}). In particular 
without the mixing matrix (\ref{zmixdef}) the correlation function of the 
operator $W_2$ with itself would not have been finite. Nor would the contact 
renormalization constants have been closed under renormalization as extra 
divergences would have appeared at two loops which could not be consistently 
renormalized. Therefore we have to treat the operator $\partial W_2$ on the 
same footing as $W_2$. 

To be more concrete we will consider the set of Green's functions 
\begin{equation}
\Sigma^L_{\mu_1 \ldots \mu_{n_L}}(p,q,r) ~=~ 
\left\langle \psi(p) \, {\cal O}^L_{\mu_1 \ldots \mu_{n_L}}(r) \, \bar{\psi}(q) 
\right\rangle 
\label{greendef}
\end{equation}
where the label $L$ denotes $S$, $V$, $T$, $W_2$ or $\partial W_2$ and the
number of Lorentz indices is $n_L$ which takes the respective values $0$, $1$,
$2$, $2$ and $2$. The three external momenta $p$, $q$ and $r$ satisfy the
energy-momentum conservation  
\begin{equation}
p ~+~ q ~+~ r ~=~ 0
\end{equation}
and we have chosen the momentum into the operator, $r$, to be the dependent
one. With this $\Sigma^L_{\mu_1 \ldots \mu_{n_L}}(p,q,r)$ will be a function of
three variables which we have chosen to be $x$, $y$ and $\mu^2$ defined by
\begin{equation}
x ~=~ \frac{p^2}{r^2} ~~~,~~~
y ~=~ \frac{q^2}{r^2} ~~~,~~~
r^2 ~=~ -~ \mu^2 ~.
\label{extconfig}
\end{equation}
where the first two are dimensionless. A related quantity which will appear in 
the final expressions for the various of the Green's function is the Gram
determinant derived from the three monenta which is given by  
\begin{equation}
\Delta_G(x,y) ~=~ 1 ~-~ 2 x ~-~ 2 y ~+~ x^2 ~-~ 2 x y ~+~ y^2 ~. 
\end{equation}
It is worth noting the connection these variables have for the earlier momentum
configurations considered in \cite{6,7,8,21,22}. The completely symmetric 
point, SMOM, is defined by $x$~$=$~$y$~$=$~$1$. However for what is now termed 
the interpolating momentum (IMOM) configuration introduced in \cite{6} there is
a subtle aspect for the mapping of the variables of (\ref{extconfig}) to those
used in \cite{22}. The main difference is that in the IMOM set up a parameter
$\omega$ was introduced with the scale of the momentum $r$ flowing in through
the operator. Therefore to make connection with the variables used here and
those of \cite{22} we note that the mapping is $x$~$\mapsto$~$\frac{1}{\omega}$
and $\mu^2$~$\mapsto$~$\omega \mu^2$. 

In order to determine (\ref{greendef}) for each operator $L$ we have 
constructed an automatic computation which evaluates the one and two loop 
Feynman graphs contributing to $\Sigma^L_{\mu_1 \ldots \mu_{n_L}}(p,q,r)$. The 
algorithm we have followed is similar to \cite{21,22} and is to decompose 
$\Sigma^L_{\mu_1 \ldots \mu_{n_L}}(p,q,r)$ via
\begin{equation}
\Sigma^L_{\mu_1 \ldots \mu_{n_L}}(p,q,r) ~=~ \sum_{k=1}^{N_L}
\Sigma^L_{(k)}(p,q) \, {\cal P}^L_{(k) \, \mu_1 \ldots \mu_{n_L} }(p,q) 
\label{decomp}
\end{equation}
into a basis of Lorentz tensors,
${\cal P}^L_{(k) \, \mu_1 \ldots \mu_{n_L} }(p,q)$, which carry the spinor
indices of the external quark fields, with an associated set of scalar 
amplitudes, $\Sigma^L_{(k)}(p,q)$. Here $N_L$ denotes the number of elements in 
the Lorentz tensor basis which are $2$, $6$, $8$ and $10$ of the respective 
sectors of (\ref{opset}). The explicit forms of the tensors in each basis are 
provided in Appendix A. Each of the amplitudes in (\ref{decomp}) is a sum of 
scalar Feynman integrals to which we can apply the Laporta algorithm, 
\cite{27}. This allows us to relate all the integrals comprising each Green's 
function through integration by parts to a set of core Feynman integrals which 
are termed masters. Their values have been determined by direct methods, 
\cite{28,29,30,31}, and we have summarized the key functions which arise in the
final expressions for $\Sigma^L_{\mu_1 \ldots \mu_{n_L}}(p,q,r)$ in Appendix B. 
To extract the integrals comprising each amplitude we use the same projection 
method of \cite{21,22} where $\Sigma^L_{\mu_1 \ldots \mu_{n_L}}(p,q,r)$ is 
multiplied by a linear combination of 
${\cal P}^L_{(k) \, \mu_1 \ldots \mu_{n_L} }(p,q)$ for each value of $k$. To 
construct the projection we have to accommodate the spinor index aspect of each
of the tensors in each basis. A systematic way to achieve this is to use a 
specific basis for all possible combinations of $\gamma$-matrices that can 
arise. These have been discussed at length in \cite{32,33,34,35,36} and are 
defined as
\begin{equation}
\Gamma_{(n)}^{\mu_1 \ldots \mu_n} ~=~ \gamma^{[\mu_1} \ldots \gamma^{\mu_n]} 
\end{equation}
where $\Gamma_{(n)}^{\mu_1 \ldots \mu_n}$, with $n$~$\geq$~$0$, are totally
antisymmetric in the Lorentz indices. There are a countably infinite number of
these matrices and they form a complete set which span spinor space in 
$d$-dimensional spacetime. This is important since we will use dimensional 
regularization to evaluate all the Feynman integrals. Clearly for an integer 
dimension the basis will truncate to a finite set but they allow one to 
systematically construct the projection matrix from the basis tensors 
${\cal P}^L_{(k) \, \mu_1 \ldots \mu_{n_L} }(p,q)$ since the
$\Gamma_{(n)}^{\mu_1 \ldots \mu_n}$ naturally partitions spinor space due to 
the property, \cite{32,33,34,35,36},
\begin{equation}
\mbox{tr} \left( \Gamma_{(m)}^{\mu_1 \ldots \mu_m}
\Gamma_{(n)}^{\nu_1 \ldots \nu_n} \right) ~ \propto ~ \delta_{mn}
I^{\mu_1 \ldots \mu_m \nu_1 \ldots \nu_n} ~.
\label{gentrace}
\end{equation}
Here $I^{\mu_1 \ldots \mu_m \nu_1 \ldots \nu_n}$ is the generalized unit matrix
on the infinite dimensional space and the trace is over the spinor indices. One
advantage of using the $\Gamma_{(n)}$-matrices is that they can only be 
contracted by external momenta which are different due to the antisymmetry 
property. For higher $n$-point functions this would allow one to construct 
tensor bases involving $\gamma$-matrices in a systematic way. In light of this 
each scalar amplitude is deduced from  
\begin{equation}
\Sigma^L_{(k)}(p,q) ~=~ {\cal M}^L_{kl}
\mbox{tr} \left( {\cal P}^{L ~\, \mu_1 \ldots \mu_{n_L}}_{(l)}(p,q) 
\Sigma^L_{\mu_1 \ldots \mu_{n_L}}(p,q,r) \right)
\end{equation}
where there is a sum over $l$. The projection matrix ${\cal M}^L_{kl}$ is 
symmetric and its entries are polynomials in $d$, $x$ and $y$. The only 
kinematic scale dependence comes through a possible overall power of $\mu^2$. 
The matrix ${\cal M}^L_{kl}$ is the inverse of the $N_L$~$\times$~$N_L$ matrix
${\cal N}^L_{kl}$ which is computed from 
\begin{equation}
{\cal N}^L_{kl} ~=~ \mbox{tr} \left( 
{\cal P}^L_{(k) \, \mu_1 \ldots \mu_{n_L}}(p,q)
{\cal P}^{L ~\, \mu_1 \ldots \mu_{n_L}}_{(l)}(p,q) \right)
\end{equation}
for each sector $L$. 

To effect the two loop computation automatically we have generated the Feynman 
graphs using {\sc Qgraf}, \cite{37}, and translated the electronic output into
the input format for the integration routine. This is written in the symbolic
manipulation language {\sc Form}, \cite{38,39}. The next step is to perform the
projection on each graph to produce a sum of scalar integrals. At this stage 
each of these carries numerators which involve scalar products of the internal 
and external momenta. These need to be simplified before the Laporta algorithm 
can be implemented. So as far as possible the scalar products are written as 
linear combinations of the propagators which in most cases reduces the number 
of propagators in the integral. In some cases the power of a propagator can 
become negative and this is regarded as what is termed an irreducible line. It 
can be accommodated within the integration by parts formalism. Therefore for 
each {\sc Qgraf} generated Feynman graph one has a set of scalar integrals 
involving positive, negative or zero powers of a set of propagators which 
describe the original topology or the original one plus irreducible ones. At 
two loops the latter could have irreducible propagators which correspond to a 
completely different topology. Again this can be accommodated within the 
Laporta formalism since the reduction to master integrals is a purely algebraic
procedure acting on integer index representations of a function constrained by 
the rules derived from the integration by parts. To achieve the reduction we 
have used the {\sc Reduze} package, \cite{40,41}, and constructed a database 
which covers all the integrals we require. From this database we have extracted
the required relations in {\sc Form} notation and included that module within 
the automatic evaluation. In terms of numbers of graphs to be computed there 
are $1$ one loop and $13$ two loop ones for $S$, $V$ and $T$. The respective
numbers for both $W_2$ and $\partial W_2$ are $3$ and $32$. The final step
after each graph has been determined to the finite part in dimensional
regularization is to carry out the renormalization in the $\MSbar$ scheme. This
is achieved using the rescaling method of \cite{42}. All graphs are computed as
a function of the bare coupling constant and gauge parameter. Then their 
renormalized counterparts are introduced via the canonical renormalization 
constant. However the operator renormalization has also to be included. For 
$S$, $V$ and $T$ this is multiplicative similar to the coupling constant while 
that for the $W_2$ sector uses the mixing matrix (\ref{zmixdef}). In each case 
this is also implemented automatically via the method of \cite{42}. For 
completeness we include the various operator $\MSbar$ renormalization group 
functions to two loops which are, \cite{23,24,43,44,45,46,47},
\begin{eqnarray}
\gamma_{S}(a) &=& -~ 3 C_F a ~+~ [ 20 T_F \Nf ~-~ 97 C_A ~-~ 6 C_F ] 
\frac{C_F a^2}{6} ~+~ O(a^3) \nonumber \\
\gamma_{V}(a) &=& O(a^3) \nonumber \\
\gamma_{T}(a) &=& C_F a ~+~ [ 257C_A ~-~ 171C_F ~-~ 52 T_F \Nf ] 
\frac{C_F a^2}{18} ~+~ O(a^3) \nonumber \\
\gamma^{W_2}_{11}(a) &=& \frac{8}{3} C_F a ~+~ \frac{1}{27} \left[ 376 C_A C_F
- 112 C_F^2 - 128 C_F T_F \Nf \right] a^2 ~+~ O(a^3) ~, \nonumber \\
\gamma^{W_2}_{12}(a) &=& -~ \frac{4}{3} C_F a ~+~ \frac{1}{27}
\left[ 56 C_F^2 - 188 C_A C_F + 64 C_F T_F \Nf \right] a^2 ~+~ O(a^3) ~, 
\nonumber \\
\gamma^{W_2}_{22}(a) &=& O(a^3) 
\end{eqnarray}
where $a$~$=$~$g^2/(16\pi^2)$ and $C_F$, $C_A$ and $T_F$ are the usual colour
group Casimirs and invariants. The anomalous dimensions $\gamma^V(a)$ and
$\gamma^{W_2}_{22}(a)$ actually vanish to all orders. The former because of the
fact that it is a physical operator and the latter as it the total derivative
of the same operator. Another reason for including the operator anomalous 
dimensions rests in a check we have on our results. Given that the finite part
of each Green's function, as will be evident, is a complicated function of the 
parameters $x$ and $y$ then the correct $\MSbar$ operator renormalization 
constants must emerge naturally in our computation. Not only that but they 
should not be $x$ or $y$ dependent which turns out to be the case. With all the
discussed ingredients we have completed the two loop evaluation of 
$\Sigma^L_{\mu_1 \ldots \mu_{n_L}}(p,q,r)$ automatically for arbitrary linear 
covariant gauge in the $\MSbar$ scheme for each of the operators in 
(\ref{opset}).

\sect{Results.}

In this section we discuss various aspects of the results and give a sense of
the properties of the various amplitudes of 
$\Sigma^L_{\mu_1 \ldots \mu_{n_L}}(p,q,r)$ for each operator $L$ of 
(\ref{opset}). We have reviewed some of the common functions of $x$ and $y$ 
which arise at one and two loops in Appendix B. They involve polylogarithms up 
to the fourth order. The expressions for the amplitudes of each of the operator
Green's functions are needless to say quite large in each case. Therefore it is
more appropriate for practical use by others to record that data in a useable
form. To achieve this we have included all the results in an attached data
file. However for completeness and to be able to give a connection to that 
notation we provide an example of one of the amplitudes. As the scalar operator
represents the most compact result the expression for the channel $1$
amplitude for this operator in the Landau gauge for the $SU(3)$ colour group
for $\Nf$~$=$~$3$ is  
\begin{eqnarray}
\left. \Sigma^S_{(1)}(p,q) \right|^{SU(3)}_{\alpha=0} &=&
-~ 1 ~+~ \left[
-~ \frac{16}{3}
+ 2 \ln(x y)
+ 2 \Phi_1 (x,y)
\right] a \nonumber \\
&& +~ \left[
60 \zeta_3
- \frac{4385}{18}
+ \frac{445}{6} \ln(x y)
- \frac{23}{3} \ln^2(x y)
- \frac{88}{9} \ln(x y) \Phi_1 (x,y)
\right. \nonumber \\
&& \left. ~~~~
+ \frac{2}{9} \ln(x y) \Phi_1 (x,y) y
- \frac{2}{9} \ln(x y) \Phi_1 (x,y) x
+ \frac{14}{3} \ln(x) \ln(y)
\right. \nonumber \\
&& \left. ~~~~
- \frac{4}{9} \ln(y) \Phi_1 (x,y) y
+ \frac{4}{9} \ln(y) \Phi_1 (x,y) x
+ \frac{1}{9} \Omega_2 \left( \frac{1}{x}, \frac{y}{x} \right) 
\right. \nonumber \\
&& \left. ~~~~
- \frac{1}{3} \Omega_2 \left( \frac{y}{x}, \frac{1}{x} \right) 
+ \frac{1}{9} \Omega_2 \left( \frac{1}{y}, \frac{x}{y} \right) 
- \frac{1}{3} \Omega_2 \left( \frac{x}{y}, \frac{1}{y} \right) 
+ \frac{2}{9} \Omega_2 (x,y)
\right. \nonumber \\
&& \left. ~~~~
+ \frac{1195}{18} \Phi_1 (x,y)
+ \frac{2}{9} \Phi_1 (x,y)^2
- \frac{2}{9} \Phi_1 (x,y)^2 y
- \frac{2}{9} \Phi_1 (x,y)^2 x
\right. \nonumber \\
&& \left. ~~~~
+ 2 \Phi_2 \left( \frac{y}{x}, \frac{1}{x} \right) \frac{1}{x} \Delta_G^{-3}
- 12 \Phi_2 \left( \frac{y}{x}, \frac{1}{x} \right) \frac{y}{x} \Delta_G^{-3}
+ 30 \Phi_2 \left( \frac{y}{x}, \frac{1}{x} \right) \frac{y^2}{x} \Delta_G^{-3}
\right. \nonumber \\
&& \left. ~~~~
- 40 \Phi_2 \left( \frac{y}{x}, \frac{1}{x} \right) \frac{y^3}{x} \Delta_G^{-3}
+ 30 \Phi_2 \left( \frac{y}{x}, \frac{1}{x} \right) \frac{y^4}{x} \Delta_G^{-3}
- 12 \Phi_2 \left( \frac{y}{x}, \frac{1}{x} \right) \frac{y^5}{x} \Delta_G^{-3}
\right. \nonumber \\
&& \left. ~~~~
+ 2 \Phi_2 \left( \frac{y}{x}, \frac{1}{x} \right) \frac{y^6}{x} \Delta_G^{-3}
- 4 \Phi_2 \left( \frac{y}{x}, \frac{1}{x} \right) \Delta_G^{-3}
- 4 \Phi_2 \left( \frac{y}{x}, \frac{1}{x} \right) \Delta_G^{-2}
\right. \nonumber \\
&& \left. ~~~~
- 4 \Phi_2 \left( \frac{y}{x}, \frac{1}{x} \right) \Delta_G^{-1}
+ 12 \Phi_2 \left( \frac{y}{x}, \frac{1}{x} \right) y \Delta_G^{-3}
+ 4 \Phi_2 \left( \frac{y}{x}, \frac{1}{x} \right) y \Delta_G^{-2}
\right. \nonumber \\
&& \left. ~~~~
- 4 \Phi_2 \left( \frac{y}{x}, \frac{1}{x} \right) y \Delta_G^{-1}
- 8 \Phi_2 \left( \frac{y}{x}, \frac{1}{x} \right) y^2 \Delta_G^{-3}
+ 4 \Phi_2 \left( \frac{y}{x}, \frac{1}{x} \right) y^2 \Delta_G^{-2}
\right. \nonumber \\
&& \left. ~~~~
- 8 \Phi_2 \left( \frac{y}{x}, \frac{1}{x} \right) y^3 \Delta_G^{-3}
- 4 \Phi_2 \left( \frac{y}{x}, \frac{1}{x} \right) y^3 \Delta_G^{-2}
+ 12 \Phi_2 \left( \frac{y}{x}, \frac{1}{x} \right) y^4 \Delta_G^{-3}
\right. \nonumber \\
&& \left. ~~~~
- 4 \Phi_2 \left( \frac{y}{x}, \frac{1}{x} \right) y^5 \Delta_G^{-3}
+ 2 \Phi_2 \left( \frac{y}{x}, \frac{1}{x} \right) x \Delta_G^{-3}
+ 2 \Phi_2 \left( \frac{y}{x}, \frac{1}{x} \right) x \Delta_G^{-2}
\right. \nonumber \\
&& \left. ~~~~
+ 2 \Phi_2 \left( \frac{y}{x}, \frac{1}{x} \right) x \Delta_G^{-1}
- 8 \Phi_2 \left( \frac{y}{x}, \frac{1}{x} \right) x y \Delta_G^{-3}
- 4 \Phi_2 \left( \frac{y}{x}, \frac{1}{x} \right) x y \Delta_G^{-2}
\right. \nonumber \\
&& \left. ~~~~
+ 12 \Phi_2 \left( \frac{y}{x}, \frac{1}{x} \right) x y^2 \Delta_G^{-3}
+ 2 \Phi_2 \left( \frac{y}{x}, \frac{1}{x} \right) x y^2 \Delta_G^{-2}
- 8 \Phi_2 \left( \frac{y}{x}, \frac{1}{x} \right) x y^3 \Delta_G^{-3}
\right. \nonumber \\
&& \left. ~~~~
+ 2 \Phi_2 \left( \frac{y}{x}, \frac{1}{x} \right) x y^4 \Delta_G^{-3}
+ 2 \Phi_2 \left( \frac{1}{y}, \frac{x}{y} \right) \frac{1}{y}
- \frac{52}{9} \Phi_2 (x,y)
\right] a^2 \nonumber \\
&& +~ O(a^3) 
\label{mass11}
\end{eqnarray}
where $\alpha$ is the gauge parameter, $\Nf$ is the number of massless quarks
and $\zeta_z$ is the Riemann zeta function. To gauge the structure of this
amplitude as a function of $x$ and $y$ we have provided a contour 
plot of it over the domain $-$~$\half$~$\leq$~$x$~$\leq$~$2$ and  
$-$~$\half$~$\leq$~$y$~$\leq$~$2$ in Figure \ref{figs} for $\Nf$~$=$~$3$ in 
the Landau gauge for the $SU(3)$ colour group. Also included in this Figure is 
the channel $2$ amplitude for comparison. Both are the one loop functions for 
the particular value of $\alpha_s$~$=$~$0.125$ with 
$\alpha_s$~$=$~$g^2/(4\pi)$. Plotting the two loop amplitudes for the same 
value of the coupling does not significantly change the qualitative behaviour 
of the amplitudes by more than a few percent.

{\begin{figure}[ht]
\begin{center}
\includegraphics[width=7.7cm,height=7.7cm]{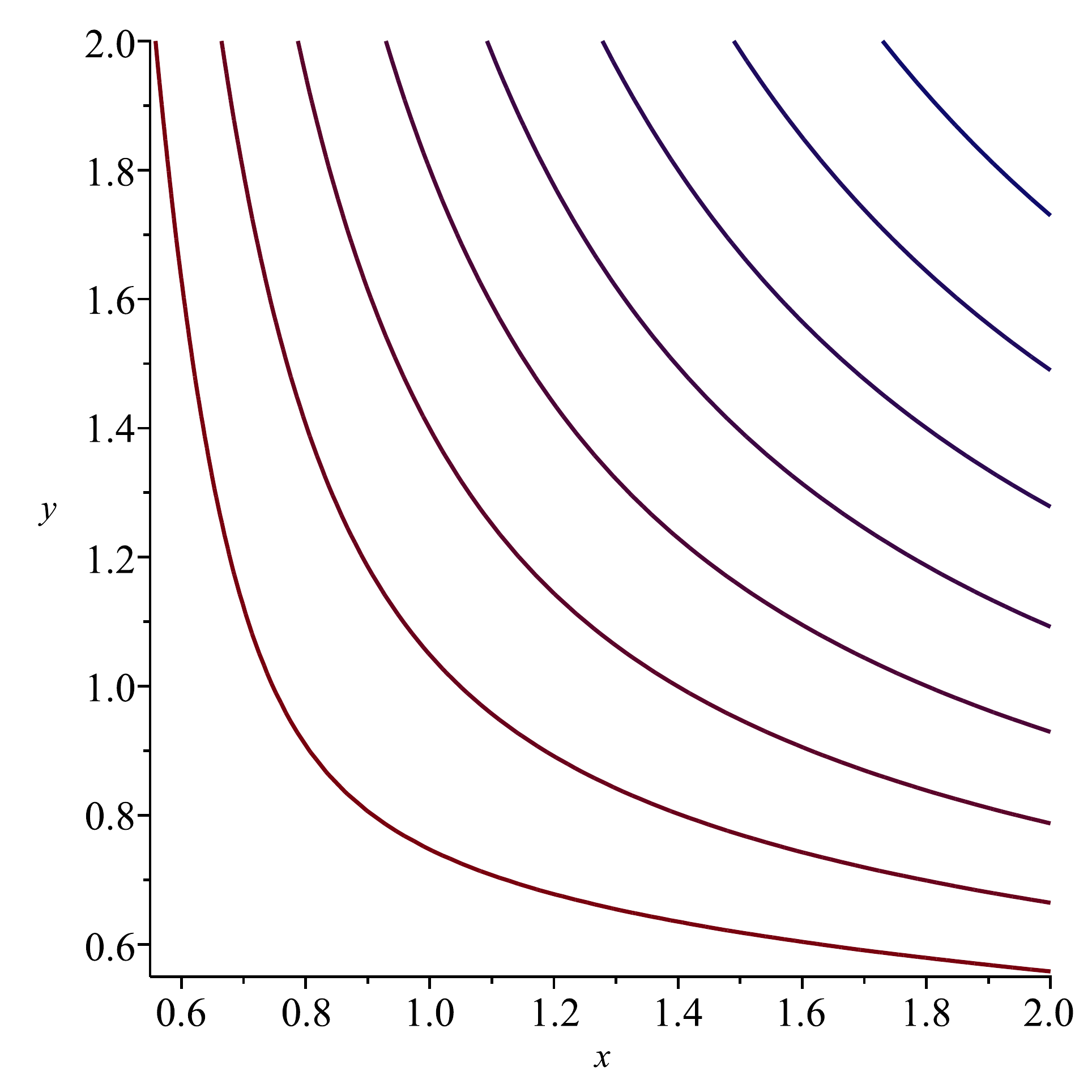}
\quad
\includegraphics[width=7.7cm,height=7.7cm]{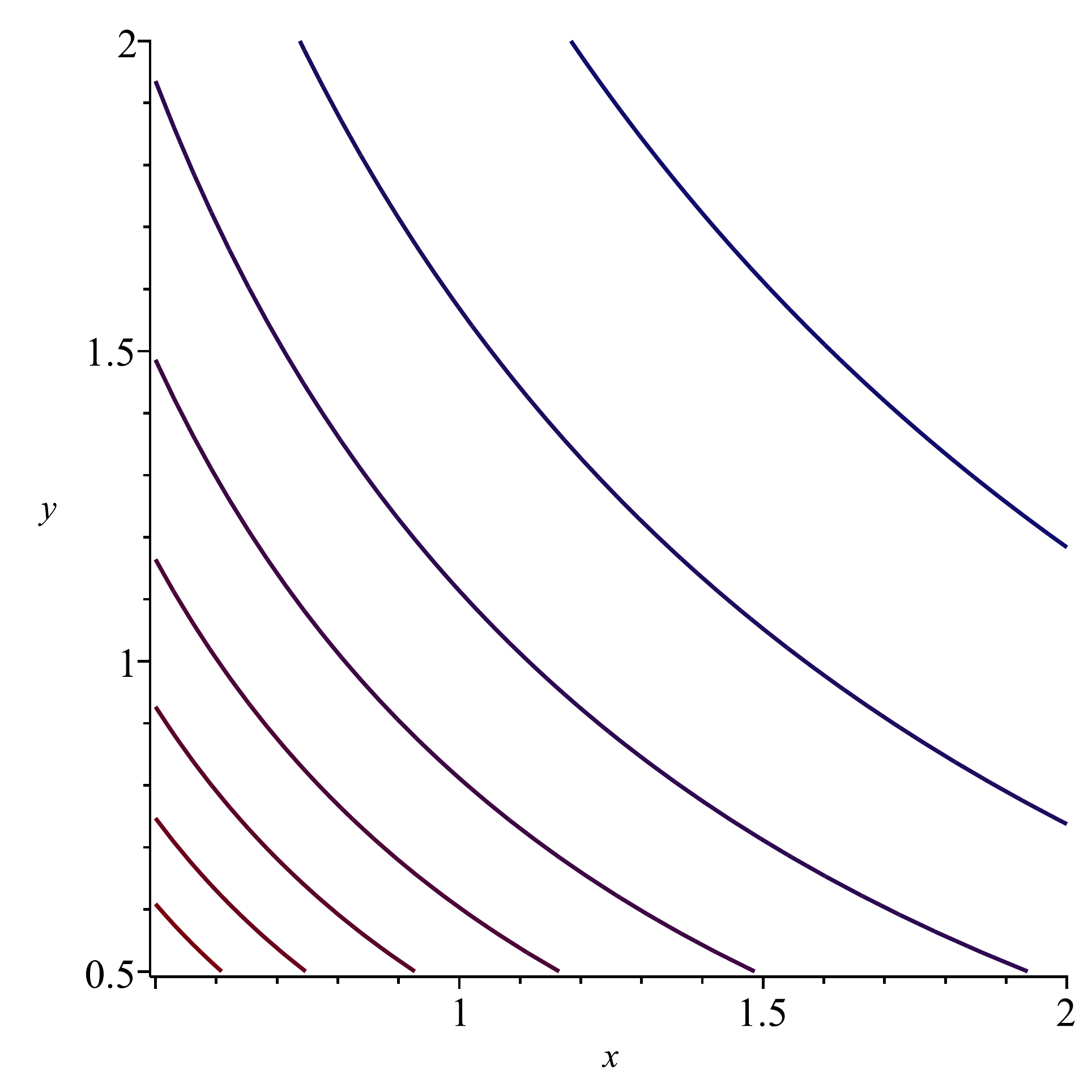}
\end{center}
\caption{One loop amplitudes $1$ (left) and $2$ (right) for $S$.}
\label{figs}
\end{figure}}

{\begin{figure}[ht]
\begin{center}
\includegraphics[width=7.7cm,height=7.7cm]{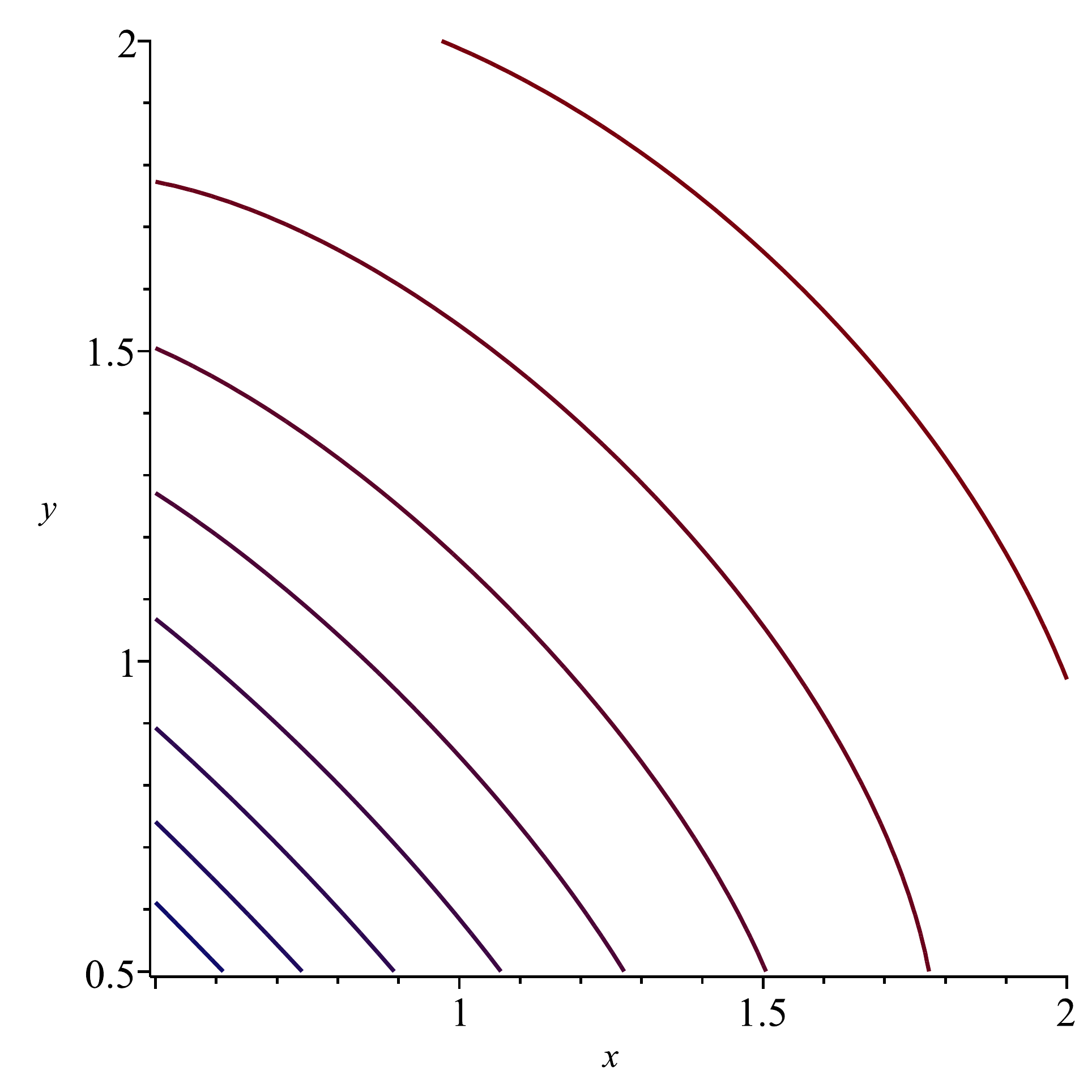}
\quad
\includegraphics[width=7.7cm,height=7.7cm]{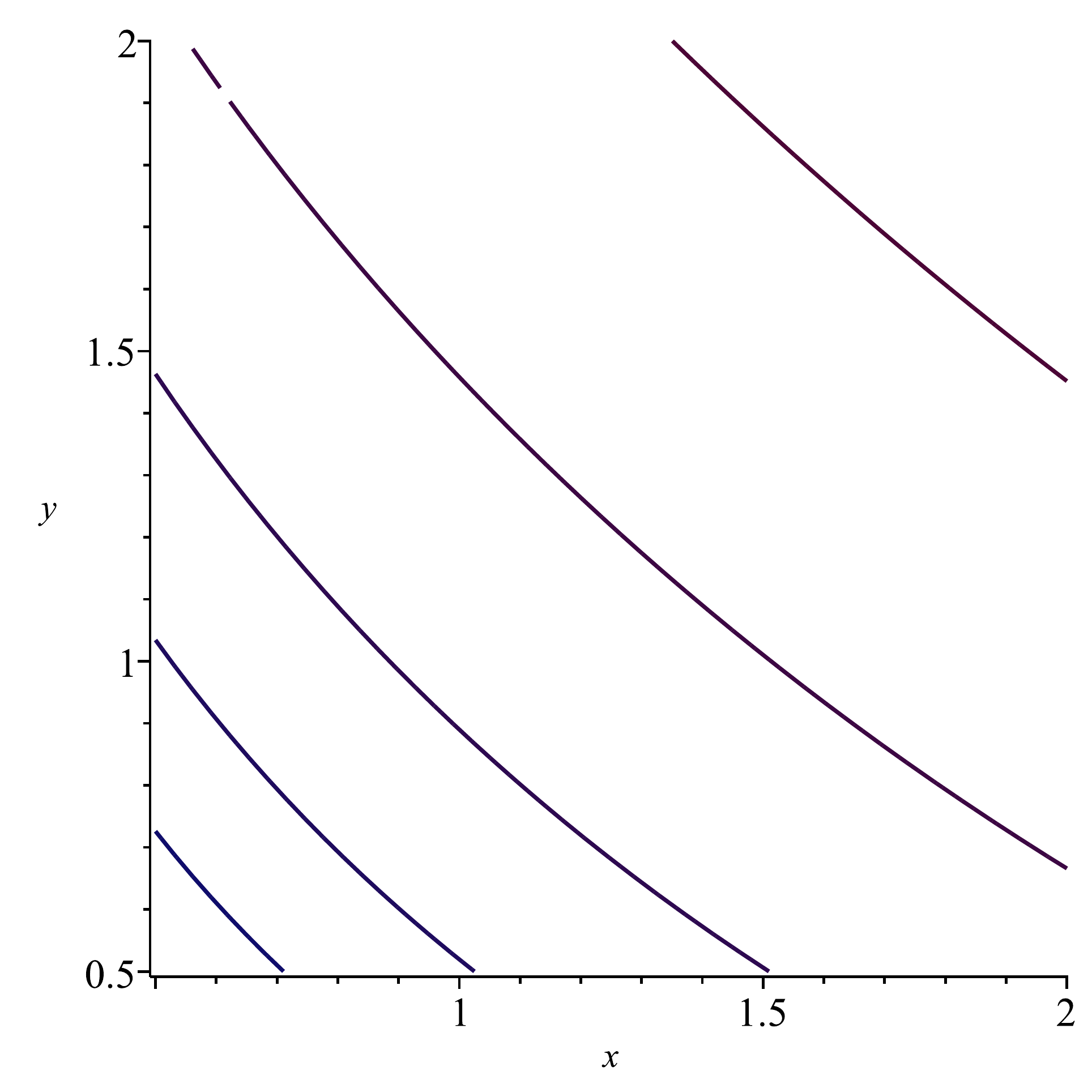}
\end{center}
\caption{One loop amplitudes $1$ (left) and $2$ (right) for $V$.}
\label{figv}
\end{figure}}

For each of the other operators the expressions for the amplitudes are formally
similar to (\ref{mass11}) but larger. To assist with appreciating the structure
of amplitudes for these other cases we have made similar contour  
plots for the same gauge, colour and flavour parameters as Figure \ref{figs} 
for the first two Lorentz channels. These are given in Figures \ref{figv}, 
\ref{figt}, \ref{figw2} and \ref{figtw2}. The behaviour of the results for $V$ 
and $T$ are similar in form. Recalling that channel $1$ for $S$, $V$ and $T$ 
contains an $O(1)$ term, but for the operator $W_2$ it is channel $2$, we see a
larger variation over the domain we have chosen for these channels compared 
with the others. Though there is an exception for $T$ which is a reflection 
that in this case channel $2$ corresponds to a different partition of the 
$\Gamma_{(n)}$-matrices. For $\partial W_2$ we have plotted channel $3$ rather
than $2$ since the latter is equivalent to the graph of channel $1$. This is
because the operator $\partial W_2$ is a total derivative and this derivative 
introduces this symmetry. Moreover the plots for channel $1$ of $V$ and
$\partial W_2$ are equivalent for similar reasons. 

{\begin{figure}[ht]
\begin{center}
\includegraphics[width=7.7cm,height=7.7cm]{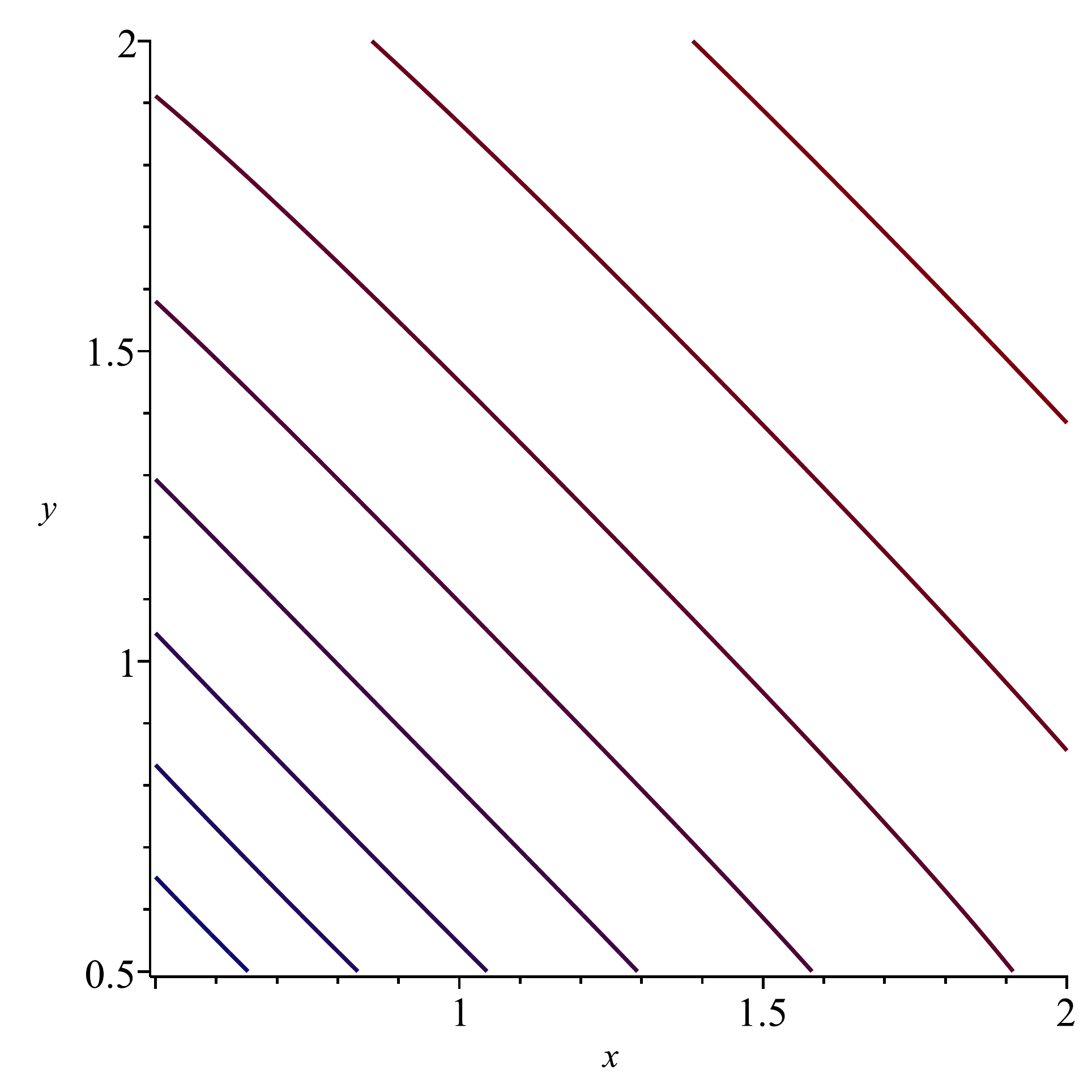}
\quad
\includegraphics[width=7.7cm,height=7.7cm]{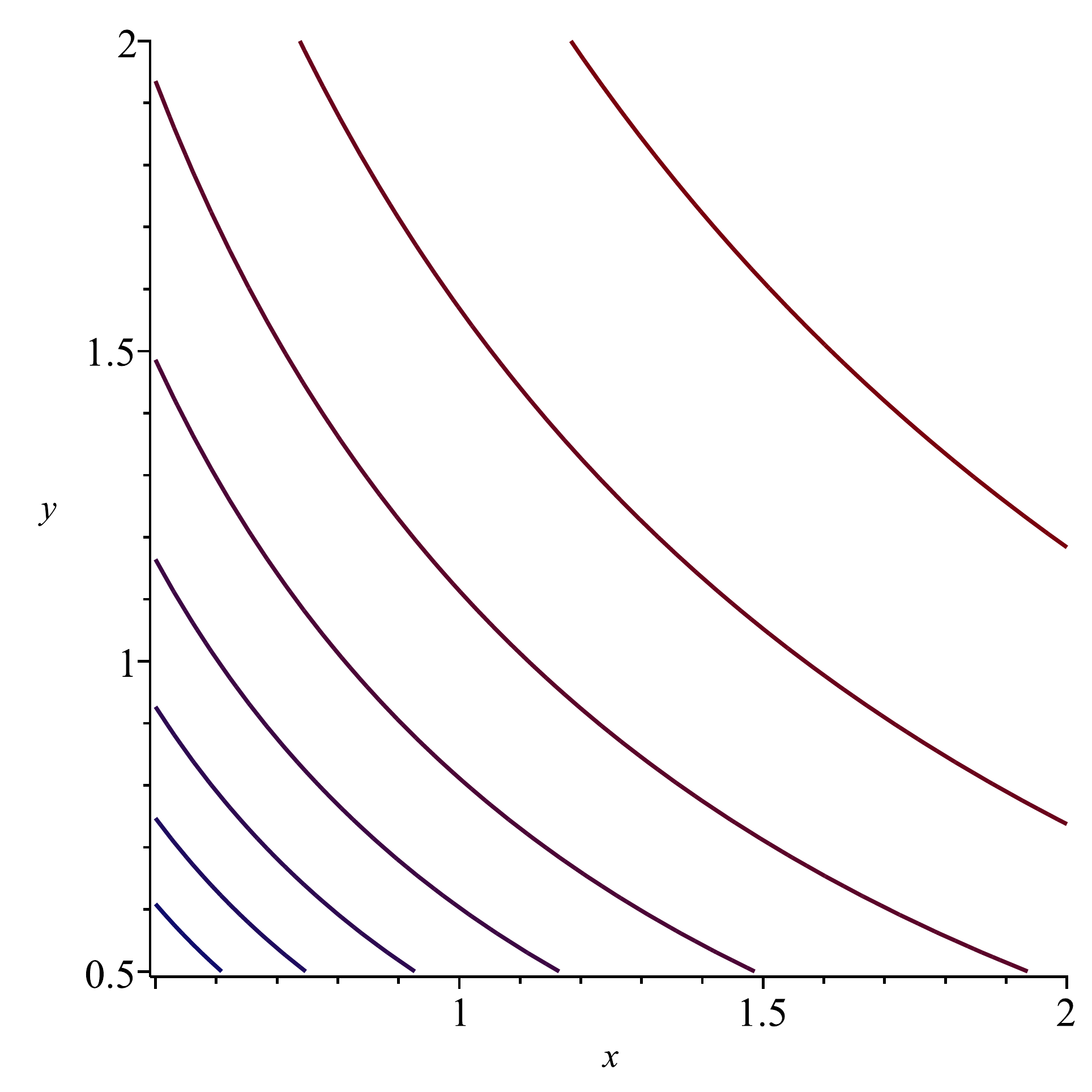}
\end{center}
\caption{One loop amplitudes $1$ (left) and $2$ (right) for $T$.}
\label{figt}
\end{figure}}

{\begin{figure}[ht]
\begin{center}
\includegraphics[width=7.7cm,height=7.7cm]{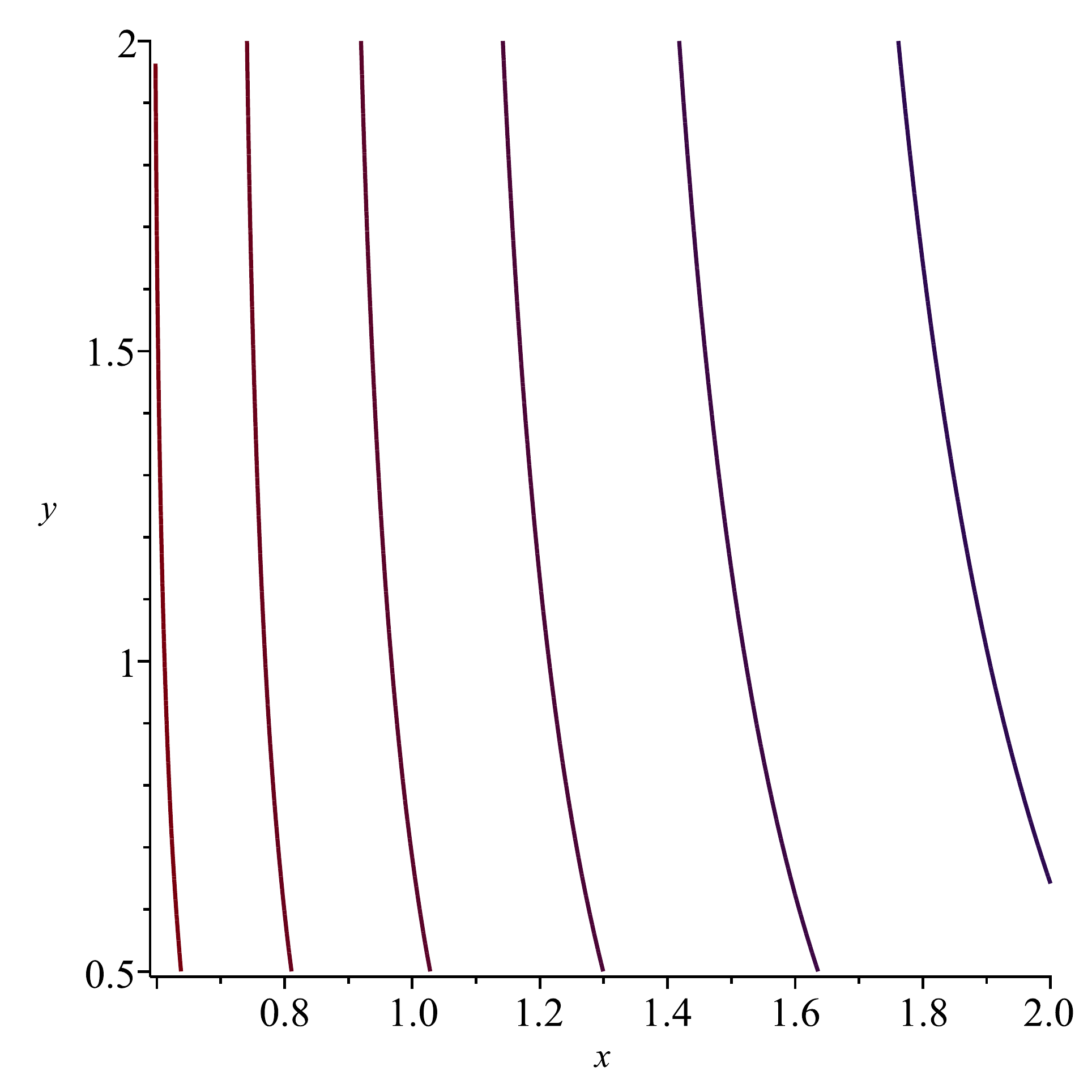}
\quad
\includegraphics[width=7.7cm,height=7.7cm]{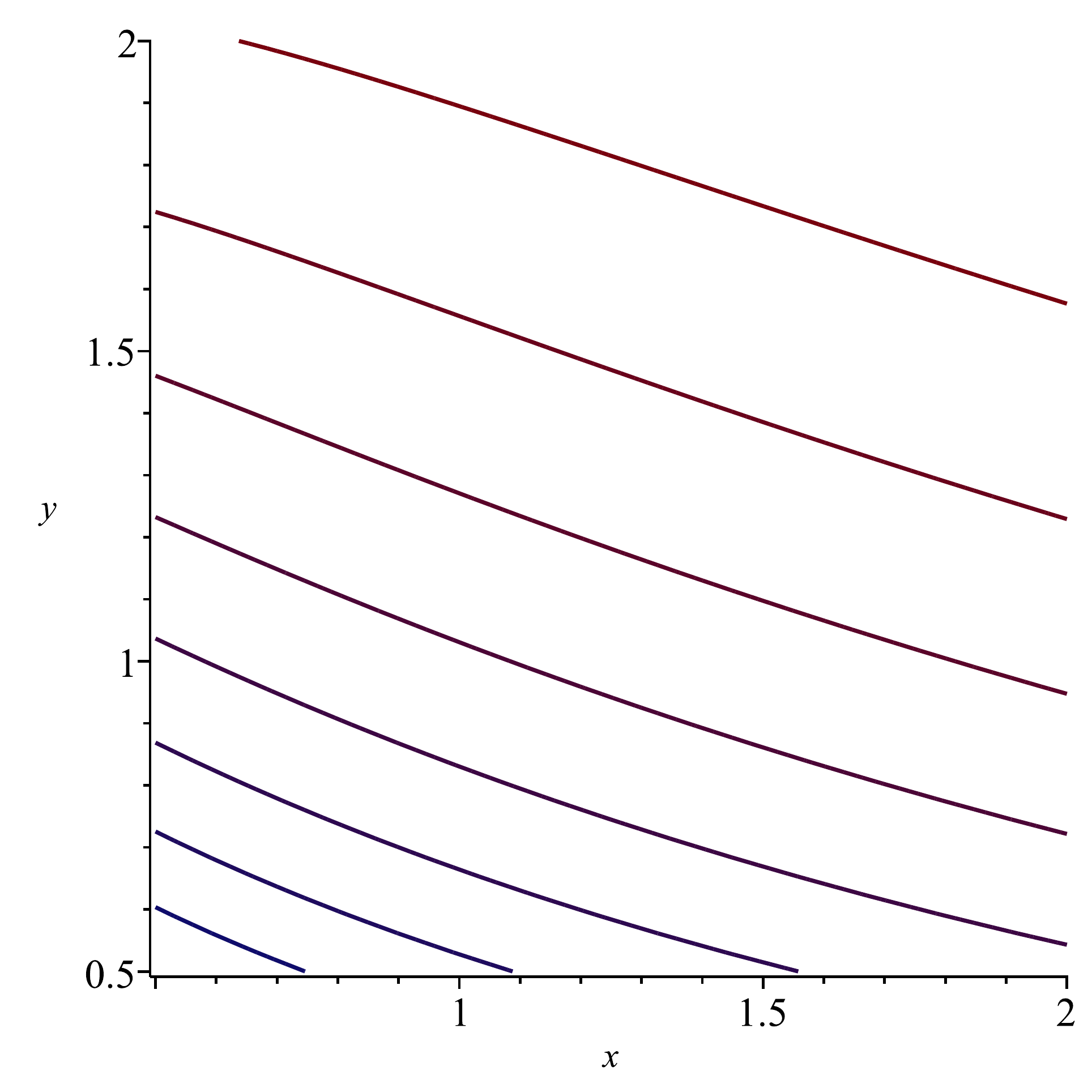}
\end{center}
\caption{One loop amplitudes $1$ (left) and $2$ (right) for $W_2$.}
\label{figw2}
\end{figure}}

{\begin{figure}[ht]
\begin{center}
\includegraphics[width=7.7cm,height=7.7cm]{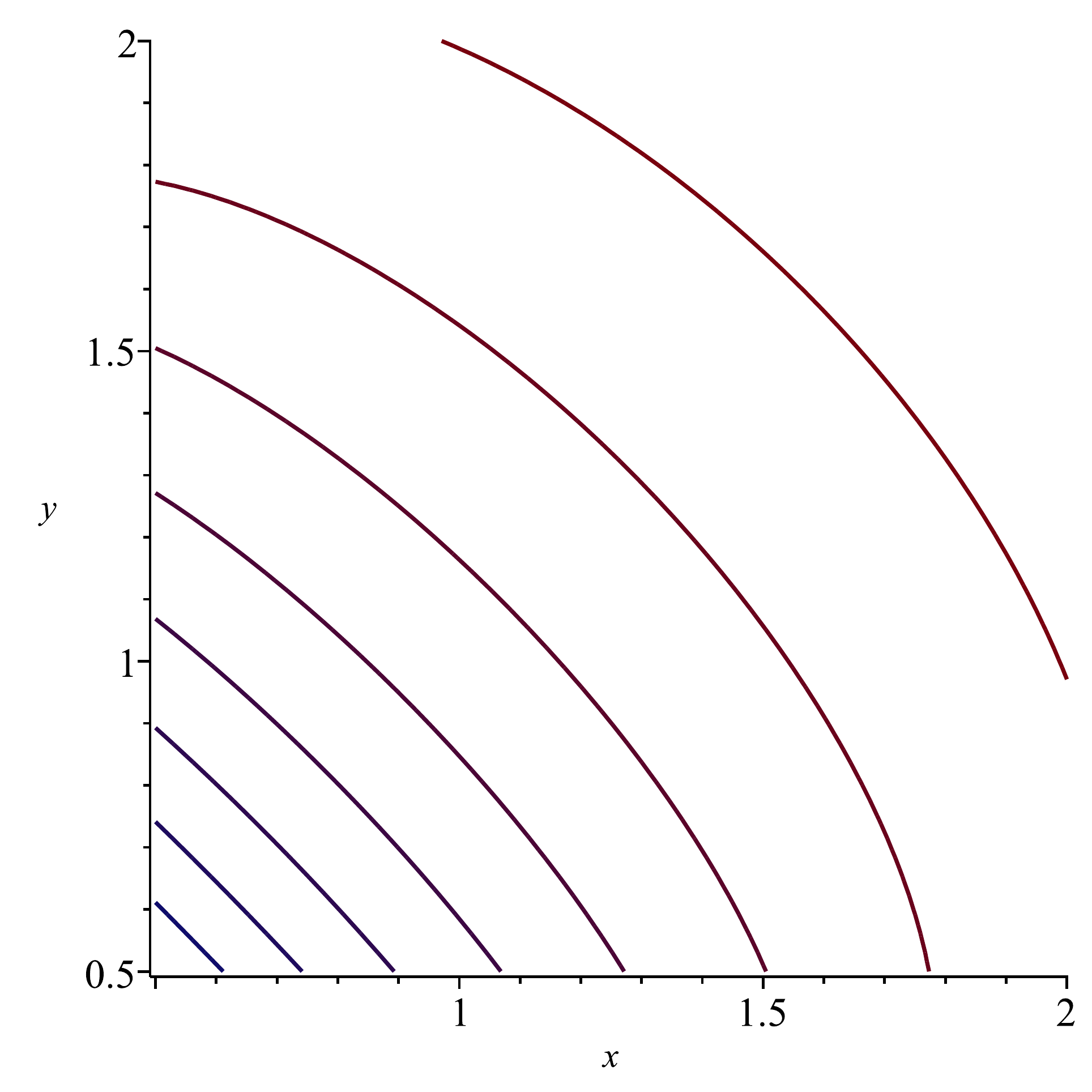}
\quad
\includegraphics[width=7.7cm,height=7.7cm]{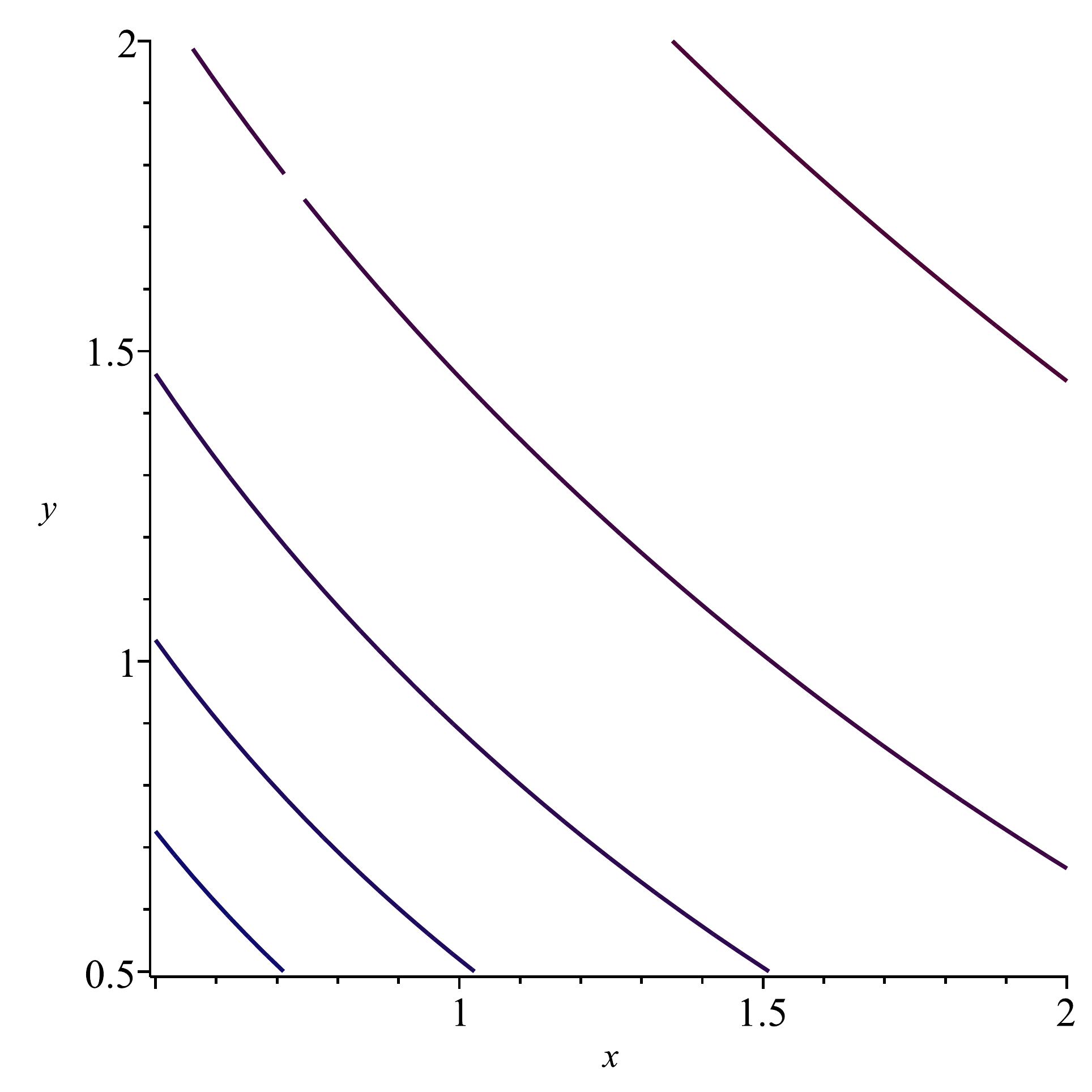}
\end{center}
\caption{One loop amplitudes $1$ (left) and $3$ (right) for $\partial W_2$.}
\label{figtw2}
\end{figure}}

One aspect of our calculation which we have checked is the generalization of
the relations between amplitudes given in \cite{21,22} are satisfied. By this 
we mean that in the original Green's function one can interchange the external 
quark and antiquark legs which implies that several of the amplitudes in the 
Lorentz decomposition are related. For the general off-shell cases this means 
that the momenta $p$ and $q$ have to be swapped in the explicit expressions. 
For completeness we note that the relations are  
\begin{equation}
\Sigma^{V}_{(2)}(p,q) ~=~
\Sigma^{V}_{(5)}(q,p) ~~~,~~~
\Sigma^{V}_{(3)}(p,q) ~=~
\Sigma^{V}_{(4)}(q,p)
\end{equation}
for the vector operator while 
\begin{equation}
\Sigma^{T}_{(3)}(p,q) ~=~
\Sigma^{T}_{(6)}(q,p) ~~~,~~~
\Sigma^{T}_{(4)}(p,q) ~=~
\Sigma^{T}_{(5)}(q,p)
\end{equation}
are the corresponding ones for the tensor case. In the $W_2$ sector due to the
asymmetry in the definition of the operator $W_2$ itself there are only
symmetry relations for the $\partial W_2$ operator. These are 
\begin{eqnarray}
\Sigma^{\partial W_2}_{(1)}(p,q) &=&
\Sigma^{\partial W_2}_{(2)}(q,p) ~~~,~~~
\Sigma^{\partial W_2}_{(3)}(p,q) ~=~
\Sigma^{\partial W_2}_{(8)}(q,p) ~~,~~ 
\Sigma^{\partial W_2}_{(4)}(p,q) ~=~
\Sigma^{\partial W_2}_{(7)}(q,p) \nonumber \\
\Sigma^{\partial W_2}_{(5)}(p,q) &=&
\Sigma^{\partial W_2}_{(6)}(q,p) ~~,~~ 
\Sigma^{\partial W_2}_{(9)}(p,q) ~=~
\Sigma^{\partial W_2}_{(10)}(q,p) ~.
\end{eqnarray}
In the case of each operator the order of the momenta arguments in the 
amplitude of the right hand side have been swapped. We have verified that each 
of the above relations for the respective operators hold to two loops for all 
$x$ and $y$. As a final check on our results we have taken the limits back to 
various results which are already known, \cite{21,22}.

\sect{Discussion.}

The computation of the Green's function (\ref{greendef}) which we have carried
out here for the operators of (\ref{opset}) in the most general off-shell 
momentum configuration completes our programme to provide their full structure 
to two loops. With the provision of different momentum values for the external 
quark fields of (\ref{greendef}) it should be possible to examine new aspects
of the dynamics of the partons of the proton for problems of current interest.
We note again that that the lattice evaluation of the pressure inside the
proton, \cite{20}, would be one physical quantity of distinct interest given
the potential to refine the comparison with the original experimental results 
of \cite{19} further. That aside there are other uses for our results. For
instance, the parton distribution functions have been considered on the lattice
in, for example, \cite{48,49,50,51,52}. Again the greater freedom to measure 
the Wilson operator Green's function in a larger set of momenta choices should 
assist with improving our knowledge of the deeper structure of the proton. The
subsequent stage to our programme will be to extend to the next loop order.
This is not a trivial task for the general momentum configuration. It would
require the expressions of the master integrals analogous to the two loop ones
of \cite{28,29,30,31}. While progress to achieve this has been made in recent 
years, \cite{53}, with the provision of the algorithm to determine the master
integrals the explicit functions are not yet known. That is the next stage in
the programme. 

\vspace{1cm}
\noindent
{\bf Acknowledgements.} This work was supported by a DFG Mercator Fellowship. 
The author thanks R. Horsley and P.E.L. Rakow for encouragement and valuable 
discussions as well as the Mathematical Physics Group at Humboldt University, 
Berlin for its hospitality where this work was initiated.

\appendix

\sect{Tensor bases and operator projection matrices.}

In this appendix we record the basis tensors for the decomposition of each
Green's functions together with the elements of each projection matrix. While
each is similar to their counterparts in previous momentum configurations,
\cite{21,22}, there are several differences in the general case where $x$ and 
$y$ are not restricted. For the scalar quark operator there are two tensors 
when there are two independent external momenta which are  
\begin{equation}
{\cal P}^{S}_{(1)}(p,q) ~=~ \Gamma_{(0)} ~~~,~~~
{\cal P}^{S}_{(2)}(p,q) ~=~ \frac{1}{\mu^2} \Gamma_{(2)}^{pq} ~.
\end{equation}
In this and the other bases the scale $\mu$ will appear in several elements to
ensure each has the same dimension. It also means that the elements of each
projection matrix have the same dimension. As the scalar operator basis 
involves different elements of the generalized $\gamma$-matrices then the 
projection matrix is diagonal due to (\ref{gentrace}) giving
\begin{equation}
{\cal M}^S ~=~ \frac{1}{4\Delta_G} \left(
\begin{array}{cc}
\Delta_G & 0 \\
0 & 4 \\
\end{array}
\right) ~.
\end{equation}
There is a similar partition for the remaining projection matrices which are
larger. 

For the vector case there are six basis elements defined as
\begin{eqnarray}
{\cal P}^{V}_{(1) \mu }(p,q) &=& \gamma_\mu ~~~,~~~
{\cal P}^{V}_{(2) \mu }(p,q) ~=~ \frac{{p}^\mu \pslash}{\mu^2} ~~~,~~~
{\cal P}^{V}_{(3) \mu }(p,q) ~=~ \frac{{p}_\mu \qslash}{\mu^2} ~, \nonumber \\
{\cal P}^{V}_{(4) \mu }(p,q) &=& \frac{{q}_\mu \pslash}{\mu^2} ~~~,~~~
{\cal P}^{V}_{(5) \mu }(p,q) ~=~ \frac{{q}_\mu \qslash}{\mu^2} ~~~,~~~
{\cal P}^{V}_{(6) \mu }(p,q) ~=~ \frac{1}{\mu^2} \Gamma_{(3) \, \mu p q}
\end{eqnarray}
where the final one will form a unit partition. However as the projection
matrix is now $6$~$\times$~$6$ but symmetric we will only list those non-zero 
elements of the upper triangle. Defining
\begin{equation}
{\cal M}^V ~=~ \frac{1}{4[d-2]\Delta_G^2}
\tilde{\cal M}^V
\end{equation}
in order to extract the overall common factor we then have 
\begin{eqnarray}
\tilde{\cal M}^V_{11} &=& [x^2 - 2 x y - 2 x + y^2 - 2 y + 1]^2 ~~,~~ 
\tilde{\cal M}^V_{12} ~=~ -~ 4 [(y - 1)^2 + x^2 - 2 (y + 1) x] y 
\nonumber \\
\tilde{\cal M}^V_{13} &=& -~ 2 [x^2 - 2 x y - 2 x + y^2 - 2 y + 1] 
[x + y - 1] \nonumber \\
\tilde{\cal M}^V_{14} &=& -~ 2 [x^2 - 2 x y - 2 x + y^2 - 2 y + 1] 
[x + y - 1] \nonumber \\
\tilde{\cal M}^V_{15} &=& -~ 4 [(y - 1)^2 + x^2 - 2 (y + 1) x] x ~~,~~ 
\tilde{\cal M}^V_{22} ~=~ 16 [d-1] y^2 \nonumber \\
\tilde{\cal M}^V_{23} &=& 8 [d-1] [x + y - 1] y ~~,~~ 
\tilde{\cal M}^V_{24} ~=~ 8 [d-1] [x + y - 1] y \nonumber \\
\tilde{\cal M}^V_{25} &=& 4 [d x^2 + 2 d x y - 2 d x + d y^2 - 2 d y + d 
- 2 x^2 + 4 x - 2 y^2 + 4 y - 2] \nonumber \\
\tilde{\cal M}^V_{33} &=& 4 [4 d x y + x^2 - 6 x y - 2 x + y^2 - 2 y + 1] ~~,~~ 
\tilde{\cal M}^V_{34} ~=~ 4 [d-1] [x + y - 1]^2 \nonumber \\
\tilde{\cal M}^V_{35} &=& 8 [y - 1 + x] [d-1] x ~~,~~ 
\tilde{\cal M}^V_{44} ~=~ 4 [4 d x y + x^2 - 6 x y - 2 x + y^2 - 2 y + 1] 
\nonumber \\
\tilde{\cal M}^V_{45} &=& 8 [y - 1 + x] [d-1] x ~~,~~ 
\tilde{\cal M}^V_{55} ~=~ 16 [d-1] x^2 \nonumber \\
\tilde{\cal M}^V_{66} &=& 4 [(y - 1)^2 + x^2 - 2 (y + 1) x] 
\end{eqnarray}
where we have not listed the zero elements outside the $\Gamma_{(1)}$ and
$\Gamma_{(3)}$ partitions. 

Following \cite{21,22} our basis for the tensor operator is 
\begin{eqnarray}
{\cal P}^{T}_{(1) \mu \nu }(p,q) &=& \Gamma_{(2) \, \mu\nu} ~~~,~~~
{\cal P}^{T}_{(2) \mu \nu }(p,q) ~=~ \frac{1}{\mu^2} \left[ p_\mu q_\nu - p_\nu
q_\mu \right] \Gamma_{(0)} ~, \nonumber \\
{\cal P}^{T}_{(3) \mu \nu }(p,q) &=& \frac{1}{\mu^2}
\left[ \Gamma_{(2) \, \mu p} p_\nu - \Gamma_{(2) \, \nu p} p_\mu
\right] ~~~,~~~
{\cal P}^{T}_{(4) \mu \nu }(p,q) ~=~ \frac{1}{\mu^2}
\left[ \Gamma_{(2) \, \mu p} q_\nu - \Gamma_{(2) \, \nu p} q_\mu
\right] ~, \nonumber \\
{\cal P}^{T}_{(5) \mu \nu }(p,q) &=& \frac{1}{\mu^2}
\left[ \Gamma_{(2) \, \mu q} p_\nu - \Gamma_{(2) \, \nu q} p_\mu
\right] ~~~,~~~
{\cal P}^{T}_{(6) \mu \nu }(p,q) ~=~ \frac{1}{\mu^2}
\left[ \Gamma_{(2) \, \mu q} q_\nu - \Gamma_{(2) \, \nu q} q_\mu
\right] ~, \nonumber \\
{\cal P}^{T}_{(7) \mu \nu }(p,q) &=& \frac{1}{\mu^4}
\left[ \Gamma_{(2) \, p q} p_\mu q_\nu - \Gamma_{(2) \, p q} p_\nu q_\mu
\right] ~~~,~~~
{\cal P}^{T}_{(8) \mu \nu }(p,q) ~=~ \frac{1}{\mu^2}
\Gamma_{(4) \, \mu \nu p q} ~.
\end{eqnarray}
To record the elements of the projection matrix we define the factorized
matrix ${\cal M}^T$ and set
\begin{equation}
{\cal M}^T ~=~ \frac{1}{4[d-2][d-3]\Delta_G^2} \tilde{\cal M}^T ~.
\end{equation}
Then
\begin{eqnarray}
\tilde{\cal M}^T_{11} &=& -~ [x^2 - 2 x y - 2 x + y^2 - 2 y + 1]^2 ~~,~~ 
\tilde{\cal M}^T_{12} ~=~ 0 \nonumber \\
\tilde{\cal M}^T_{13} &=& 4 [(y - 1)^2 + x^2 - 2 (y + 1) x] y ~~,~~ 
\tilde{\cal M}^T_{14} ~=~ 2 [x^2 - 2 x y - 2 x + y^2 - 2 y + 1] 
[x + y - 1] \nonumber \\
\tilde{\cal M}^T_{15} &=& 2 [x^2 - 2 x y - 2 x + y^2 - 2 y + 1] 
[x + y - 1] ~~,~~ 
\tilde{\cal M}^T_{16} ~=~ 4 [(y - 1)^2 + x^2 - 2 (y + 1) x] x \nonumber \\ 
\tilde{\cal M}^T_{17} &=& 4 [(y - 1)^2 + x^2 - 2 (y + 1) x] \nonumber \\
\tilde{\cal M}^T_{22} &=& -~ 2 [(y - 1)^2 + x^2 - 2 (y + 1) x] [d-2] [d - 3]
\nonumber \\
\tilde{\cal M}^T_{23} &=&  
\tilde{\cal M}^T_{24} ~=~ 
\tilde{\cal M}^T_{25} ~=~ 
\tilde{\cal M}^T_{26} ~=~ 
\tilde{\cal M}^T_{27} ~=~ 0 ~~,~~ 
\tilde{\cal M}^T_{33} ~=~ -~ 8 [d-1] y^2 \nonumber \\
\tilde{\cal M}^T_{34} &=& -~ 4 [d-1] [x + y - 1] y ~~,~~ 
\tilde{\cal M}^T_{35} ~=~ -~ 4 [d-1] [x + y - 1] y \nonumber \\ 
\tilde{\cal M}^T_{36} &=& -~ 2 [d x^2 + 2 d x y - 2 d x + d y^2 - 2 d y 
+ d - 3 x^2 + 2 x y + 6 x - 3 y^2 + 6 y - 3] \nonumber \\
\tilde{\cal M}^T_{37} &=& -~ 8 [d-1] y ~~,~~ 
\tilde{\cal M}^T_{44} ~=~ -~ 4 [2 d x y + x^2 - 4 x y - 2 x + y^2 - 2 y
+ 1] \nonumber \\
\tilde{\cal M}^T_{45} &=& -~ 2 [d-1] [x + y - 1]^2 ~~,~~
\tilde{\cal M}^T_{46} ~=~ -~ 4 [y - 1 + x] [d-1] x \nonumber \\
\tilde{\cal M}^T_{47} &=& -~ 4 [y - 1 + x] [d-1] ~~,~~ 
\tilde{\cal M}^T_{55} ~=~ -~ 4 [2 d x y + x^2 - 4 x y - 2 x + y^2 - 2 y 
+ 1] \nonumber \\
\tilde{\cal M}^T_{56} &=& -~ 4 [y - 1 + x] [d-1] x ~~,~~ 
\tilde{\cal M}^T_{57} ~=~ -~ 4 [y - 1 + x] [d-1] ~~,~~ 
\tilde{\cal M}^T_{66} ~=~ -~ 8 [d-1] x^2 \nonumber \\
\tilde{\cal M}^T_{67} &=& -~ 8 [d-1] x ~~,~~
\tilde{\cal M}^T_{77} ~=~ -~ 8 [d-1] [d-2] \nonumber \\
\tilde{\cal M}^T_{88} &=& -~ 4 [(y - 1)^2 + x^2 - 2 (y + 1) x] 
\end{eqnarray}
are the upper triangle entries in the symmetric matrix. 

The situation for the final operator $W_2$ is slightly different from the
previous ones. In this we have chosen to define the basis in such a way that
each Lorentz tensor is symmetric and traceless. While there is no a priori 
reason for doing so it results in some of our basis elements having $x$ and $y$
dependence unlike the derivative free operators. So the basis tensors formally
differ from those of \cite{21,22}. However they equate to the latter in the
respective limits. Our choice here is
\begin{eqnarray}
{\cal P}^{W_2}_{(1) \mu \nu }(p,q) &=& \gamma_\mu p_\nu + \gamma_\nu p_\mu
- \frac{2}{d} \pslash \eta_{\mu\nu} ~~,~~
{\cal P}^{W_2}_{(2) \mu \nu }(p,q) ~=~ \gamma_\mu q_\nu + \gamma_\nu q_\mu
- \frac{2}{d} \qslash \eta_{\mu\nu} \nonumber \\
{\cal P}^{W_2}_{(3) \mu \nu }(p,q) &=& \pslash \left[
\frac{1}{\mu^2} p_\mu p_\nu + \frac{x}{d} \eta_{\mu\nu} \right] ~,~
{\cal P}^{W_2}_{(4) \mu \nu }(p,q) ~=~ \pslash \left[
\frac{1}{\mu^2} p_\mu q_\nu + \frac{1}{\mu^2} q_\mu p_\nu
+ \frac{[1-x-y]}{d} \eta_{\mu\nu} \right] \nonumber \\
{\cal P}^{W_2}_{(5) \mu \nu }(p,q) &=& \pslash \left[
\frac{1}{\mu^2} q_\mu q_\nu + \frac{y}{d} \eta_{\mu\nu} \right] ~,~
{\cal P}^{W_2}_{(6) \mu \nu }(p,q) ~=~ \qslash \left[
\frac{1}{\mu^2} p_\mu p_\nu + \frac{x}{d} \eta_{\mu\nu} \right] \nonumber \\
{\cal P}^{W_2}_{(7) \mu \nu }(p,q) &=& \qslash \left[
\frac{1}{\mu^2} p_\mu q_\nu + \frac{1}{\mu^2} q_\mu p_\nu
+\frac{[1-x-y]}{d} \eta_{\mu\nu} \right] ~,~
{\cal P}^{W_2}_{(8) \mu \nu }(p,q) ~=~ \qslash \left[
\frac{1}{\mu^2} q_\mu q_\nu + \frac{y}{d} \eta_{\mu\nu} \right] \nonumber \\
{\cal P}^{W_2}_{(9) \mu \nu }(p,q) &=& \frac{1}{\mu^2} \left[
\Gamma_{(3) \, \mu p q } p_\nu + \Gamma_{(3) \, \nu p q } p_\mu \right] ~,~ 
{\cal P}^{W_2}_{(10) \mu \nu }(p,q) ~=~ \frac{1}{\mu^2} \left[
\Gamma_{(3) \, \mu p q } q_\nu + \Gamma_{(3) \, \nu p q } q_\mu \right] ~.
\end{eqnarray}
This partitions the projection matrix into an $8$~$\times$~$8$ sub-matrix for
the $\Gamma_{(1)}$-matrices and $2$~$\times$~$2$ for the $\Gamma_{(3)}$
sector. Defining  
\begin{equation}
{\cal M}^{W_2} ~=~ \frac{1}{4[d-2]^2\Delta_G^3\mu^2} \tilde{\cal M}^{W_2}
\end{equation}
where the factor includes $\mu^2$ since the elements of the tensor basis each
have an odd number of external momenta. The non-zero elements of the upper 
triangle of each sub-matrix of the two symmetric partitions of 
$\tilde{\cal M}^{W_2}$ are
\begin{eqnarray}
\tilde{\cal M}^{W_2}_{11} &=& 2 [d - 2] [x^2 - 2 x y - 2 x + y^2 - 2 y + 1]^2 y \nonumber \\
\tilde{\cal M}^{W_2}_{12} &=& [d - 2] [x^2 - 2 x y - 2 x + y^2 - 2 y + 1]^2 [x + y - 1] \nonumber \\
\tilde{\cal M}^{W_2}_{13} &=& -~ 16 [[y - 1]^2 + x^2 - 2 [y + 1] x] [d - 2] y^2 \nonumber \\
\tilde{\cal M}^{W_2}_{14} &=& -~ 8 [d - 2] [x^2 - 2 x y - 2 x + y^2 - 2 y + 1] [x + y - 1] y \nonumber \\
\tilde{\cal M}^{W_2}_{15} &=& -~ 4 [d - 2] [x^2 - 2 x y - 2 x + y^2 - 2 y + 1] [x + y - 1]^2 \nonumber \\
\tilde{\cal M}^{W_2}_{16} &=& -~ 8 [d - 2] [x^2 - 2 x y - 2 x + y^2 - 2 y + 1] [x + y - 1] y \nonumber \\
\tilde{\cal M}^{W_2}_{17} &=& -~ 2 [d - 2] [x^2 + 6 x y - 2 x + y^2 - 2 y + 1] [x^2 - 2 x y - 2 x + y^2 - 2 y + 1] \nonumber \\
\tilde{\cal M}^{W_2}_{18} &=& -~ 8 [d - 2] [x^2 - 2 x y - 2 x + y^2 - 2 y + 1] [x + y - 1] x \nonumber \\
\tilde{\cal M}^{W_2}_{22} &=& 2 [d - 2] [x^2 - 2 x y - 2 x + y^2 - 2 y + 1]^2 x \nonumber \\ 
\tilde{\cal M}^{W_2}_{23} &=& -~ 8 [d - 2] [x^2 - 2 x y - 2 x + y^2 - 2 y + 1] [x + y - 1] y \nonumber \\
\tilde{\cal M}^{W_2}_{24} &=& -~ 2 [d - 2] [x^2 + 6 x y - 2 x + y^2 - 2 y + 1] [x^2 - 2 x y - 2 x + y^2 - 2 y + 1] \nonumber \\
\tilde{\cal M}^{W_2}_{25} &=& -~ 8 [d - 2] [x^2 - 2 x y - 2 x + y^2 - 2 y + 1] [x + y - 1] x \nonumber \\
\tilde{\cal M}^{W_2}_{26} &=& -~ 4 [d - 2] [x^2 - 2 x y - 2 x + y^2 - 2 y + 1] [x + y - 1]^2 \nonumber \\
\tilde{\cal M}^{W_2}_{27} &=& -~ 8 [d - 2] [x^2 - 2 x y - 2 x + y^2 - 2 y + 1] [x + y - 1] x \nonumber \\
\tilde{\cal M}^{W_2}_{28} &=& -~ 16 [[y - 1]^2 + x^2 - 2 [y + 1] x] [d - 2] x^2
~~,~~
\tilde{\cal M}^{W_2}_{33} ~=~ 64 [d + 1] [d - 2] y^3 \nonumber \\ 
\tilde{\cal M}^{W_2}_{34} &=& 32 [d + 1] [d - 2] [x + y - 1] y^2 \nonumber \\
\tilde{\cal M}^{W_2}_{35} &=& 16 [d x^2 + 2 d x y - 2 d x + d y^2 - 2 d y + d + 4 x y] [d - 2] y \nonumber \\
\tilde{\cal M}^{W_2}_{36} &=& 32 [d + 1] [d - 2] [x + y - 1] y^2 \nonumber \\ 
\tilde{\cal M}^{W_2}_{37} &=& 16 [d x^2 + 2 d x y - 2 d x + d y^2 - 2 d y + d + 4 x y] [d - 2] y \nonumber \\
\tilde{\cal M}^{W_2}_{38} &=& 8 [d (x+y-1)^2 + 8 x y + 4 x - 2 y^2 + 4 y - 2] [d - 2] [x + y - 1] \nonumber \\
\tilde{\cal M}^{W_2}_{44} &=& 8 [d x^2 + 6 d x y - 2 d x + d y^2 - 2 d y + d + 3 x^2 + 2 x y - 6 x + 3 y^2 - 6 y + 3] [d - 2] y \nonumber \\
\tilde{\cal M}^{W_2}_{45} &=& 8 [4 d x y + x^2 + 2 x y - 2 x + y^2 - 2 y + 1] [d - 2] [x + y - 1] \nonumber \\
\tilde{\cal M}^{W_2}_{46} &=& 16 [d + 1] [d - 2] [x + y - 1]^2 y \nonumber \\
\tilde{\cal M}^{W_2}_{47} &=& 4 [d + 1] [d - 2] [x^2 + 6 x y - 2 x + y^2 - 2 y + 1] [x + y - 1] \nonumber \\
\tilde{\cal M}^{W_2}_{48} &=& 16 [d x^2 + 2 d x y - 2 d x + d y^2 - 2 d y + d + 4 x y] [d - 2] x \nonumber \\
\tilde{\cal M}^{W_2}_{55} &=& 32 [2 d x y + x^2 - 2 x + y^2 - 2 y + 1] [d - 2] x \nonumber \\
\tilde{\cal M}^{W_2}_{56} &=& 8 [d x^2 + 2 d x y - 2 d x + d y^2 - 2 d y + d + 4 x y] [d - 2] [x + y - 1] \nonumber \\
\tilde{\cal M}^{W_2}_{57} &=& 16 [d + 1] [d - 2] [x + y - 1]^2 x ~~,~~ 
\tilde{\cal M}^{W_2}_{58} ~=~ 32 [d + 1] [d - 2] [x + y - 1] x^2 \nonumber \\
\tilde{\cal M}^{W_2}_{66} &=& 32 [2 d x y + x^2 - 2 x + y^2 - 2 y + 1] [d - 2] y \nonumber \\
\tilde{\cal M}^{W_2}_{67} &=& 8 [4 d x y + x^2 + 2 x y - 2 x + y^2 - 2 y + 1] [d - 2] [x + y - 1] \nonumber \\
\tilde{\cal M}^{W_2}_{68} &=& 16 [d x^2 + 2 d x y - 2 d x + d y^2 - 2 d y + d + 4 x y] [d - 2] x \nonumber \\
\tilde{\cal M}^{W_2}_{77} &=& 8 [d x^2 + 6 d x y - 2 d x + d y^2 - 2 d y + d + 3 x^2 + 2 x y - 6 x + 3 y^2 - 6 y + 3] [d - 2] x \nonumber \\
\tilde{\cal M}^{W_2}_{78} &=& 32 [d + 1] [d - 2] [x + y - 1] x^2 ~~,~~
\tilde{\cal M}^{W_2}_{88} ~=~ 64 [d + 1] [d - 2] x^3 \nonumber \\
\tilde{\cal M}^{W_2}_{99} &=& 8 \Delta_G^2 [d - 2] y ~~,~~ 
\tilde{\cal M}^{W_2}_{910} ~=~ 4 [d - 2] \Delta_G [x + y - 1] \nonumber \\
\tilde{\cal M}^{W_2}_{1010} &=& 8 \Delta_G [d - 2] x ~.
\end{eqnarray}
 
\sect{Basic integrals.}

In the final expressions for the operator Green's functions several core
functions arise which are combinations of the polylogarithm function
$\mbox{Li}_n(z)$. We record them here for completeness. The main function at 
one loop is 
\begin{equation}
\Phi_1(x,y) ~=~ \frac{1}{\lambda} \left[ 2 \mbox{Li}_2(-\rho x)
+ 2 \mbox{Li}_2(-\rho y)
+ \ln \left( \frac{y}{x} \right)
\ln \left( \frac{(1+\rho y)}{(1+\rho x)} \right)
+ \ln(\rho x) \ln(\rho y) + \frac{\pi^2}{3} \right]
\end{equation}
where $\lambda(x,y)$ and $\rho(x,y)$ are given by, \cite{29,30}, 
\begin{equation}
\lambda(x,y) ~=~ \sqrt{\Delta_G} ~~~,~~~
\rho(x,y) ~=~ \frac{2}{[1-x-y+\lambda(x,y)]} 
\end{equation}
and throughout this section $x$ and $y$ are variables in general not to be
confused with the kinematic ones of (\ref{extconfig}). However the triangle 
graph where $\Phi_1(x,y)$ arises has an $O(\epsilon)$ correction which cannot 
be neglected a priori for the two loop evaluation. It is given by, 
\cite{29,30},
\begin{eqnarray}
\Psi_1(x,y) &=& -~ \frac{1}{\lambda} \left[ 
4 \mbox{Li}_3 \left( - \frac{\rho x(1+\rho y)}{(1-\rho^2 xy)} \right) 
+ 4 \mbox{Li}_3 \left( - \frac{\rho y(1+\rho x)}{(1-\rho^2 xy)} \right) 
- 4 \mbox{Li}_3 \left( - \frac{xy \rho^2}{(1-\rho^2 xy)} \right) 
\right. \nonumber \\
&& \left. ~~~~~~~
+ 2 \mbox{Li}_3 \left( \frac{x \rho(1+\rho y)}{(1+\rho x)} \right) 
+ 2 \mbox{Li}_3 \left( \frac{y \rho(1+\rho x)}{(1+\rho y)} \right) 
- 2 \mbox{Li}_3 ( \rho^2 xy ) 
- 2 \zeta_3
\right. \nonumber \\
&& \left. ~~~~~~~
- 2 \ln (y) \mbox{Li}_2 \left( \frac{x \rho(1+\rho y)}{(1+\rho x)} \right) 
- 2 \ln (x) \mbox{Li}_2 \left( \frac{y \rho(1+\rho x)}{(1+\rho y)} \right) 
- \frac{2}{3} \ln^3 \left( 1-\rho^2 xy \right)
\right. \nonumber \\
&& \left. ~~~~~~~
+ \frac{2}{3} \ln^3 \left( 1+\rho x \right)
+ \frac{2}{3} \ln^3 \left( 1+\rho y \right)
+ 2 \ln(\rho) \ln^2 \left( 1-\rho^2 x y \right)
\right. \nonumber \\
&& \left. ~~~~~~~
- 2 \ln(1-\rho^2 xy ) \left[ \ln(\rho x) \ln(\rho y)
+ \ln \left( \frac{y}{x} \right)
\ln \left( \frac{(1+\rho y)}{(1+\rho x)} \right)
\right. \right. \nonumber \\
&& \left. \left. ~~~~~~~~~~~~~~~~~~~~~~~~~~~~~~
+ 2 \ln(1+\rho x) \ln(1+\rho y) + \frac{\pi^2}{3} \right] 
\right. \nonumber \\
&& \left. ~~~~~~~
+ \frac{1}{2} \ln \left( xy \rho^2 \right) \left[ \ln(\rho x) \ln(\rho y)
+ \ln \left( \frac{y}{x} \right)
\ln \left( \frac{(1+\rho y)}{(1+\rho x)} \right)
- \ln^2 \left( \frac{(1+ \rho x)}{(1+\rho y)} \right) 
\right. \right. \nonumber \\
&& \left. \left. ~~~~~~~~~~~~~~~~~~~~~~~~~~~
+ \frac{2\pi^2}{3} 
\right] \right] ~. 
\end{eqnarray}
At the next loop order there are two key functions in the two loop master
integrals. These are, \cite{28,29},
\begin{eqnarray}
\Phi_2(x,y) &=& \frac{1}{\lambda} \left[ 6 \mbox{Li}_4(-\rho x)
+ 6 \mbox{Li}_4(-\rho y)
+ 3 \ln \left( \frac{y}{x} \right)
\left[ \mbox{Li}_3(-\rho x) - \mbox{Li}_3(-\rho y) \right] \right. \nonumber \\
&& \left. ~~~
+ \frac{1}{2} \ln^2 \left( \frac{y}{x} \right)
\left[ \mbox{Li}_2(-\rho x) + \mbox{Li}_2(-\rho y) \right] 
+ \frac{1}{4} \ln^2(\rho x) \ln^2(\rho y) 
\right. \nonumber \\
&& \left. ~~~
+ \frac{\pi^2}{2} \ln(\rho x) \ln(\rho y) 
+ \frac{\pi^2}{12} \ln^2 \left( \frac{y}{x} \right)
+ \frac{7\pi^4}{60} \right] 
\end{eqnarray}
and
\begin{eqnarray}
\Omega_2(x,y) &=& 6 \mbox{Li}_3(-\rho x) + 6 \mbox{Li}_3(-\rho y)
+ 3 \ln \left( \frac{y}{x} \right)
\left[ \mbox{Li}_2(-\rho x) - \mbox{Li}_2(-\rho y) \right] \nonumber \\
&& -~ \frac{1}{2} \ln^2 \left( \frac{y}{x} \right)
\left[ \ln(1+\rho x) + \ln(1+\rho y) \right] 
\nonumber \\
&& +~ \frac{1}{2} \left[ \pi^2 + \ln(\rho x) \ln(\rho y) \right]
\left[ \ln(\rho x) + \ln(\rho y) \right] ~. 
\end{eqnarray}
These functions are related to cyclotomic polylogarithms, \cite{54}.

\end{document}